\documentclass[10pt,onecolumn]{article}

\usepackage{amsmath}
\usepackage{amssymb}
\usepackage{graphicx}
\usepackage{morefloats}
\usepackage{xcolor}

\title{\textbf{Reconstructing topological properties of complex networks using the fitness model}}

\author{Giulio Cimini$^{1}$, Tiziano Squartini$^{1}$, Nicol\`o Musmeci$^2$, Michelangelo Puliga$^3$,\\Andrea Gabrielli$^{1,3}$, Diego Garlaschelli$^{4}$, Stefano Battiston$^5$, Guido Caldarelli$^3$\\ \ \\
$^1$ Institute for Complex Systems (ISC-CNR) UoS ``Sapienza'' University of Rome (Italy)\\[1mm]
$^2$ King's College, London (United Kingdom)\\[1mm]
$^3$ IMT Institute for Advanced Studies, Lucca (Italy)\\[1mm]
$^4$ Lorentz Institute for Theoretical Physics, University of Leiden (Netherlands)\\[1mm]
$^5$ Department of Banking and Finance, University of Zurich (Switzerland)}
\date{}

\begin{document}
\maketitle

\begin{abstract}
A major problem in the study of complex socioeconomic systems is represented by privacy issues---that can put severe limitations on the amount of accessible information, 
forcing to build models on the basis of incomplete knowledge. 
In this paper we investigate a novel method to reconstruct global topological properties of a complex network starting from limited information.
This method uses the knowledge of an intrinsic property of the nodes (indicated as {\em fitness}), and the number of connections of only a limited subset of nodes, 
in order to generate an ensemble of {\em exponential random graphs} that are representative of the real systems and that can be used to estimate its topological properties. 
Here we focus in particular on reconstructing the most basic properties that are commonly used to describe a network: density of links, assortativity, clustering. 
We test the method on both benchmark synthetic networks and real economic and financial systems, finding a remarkable robustness with respect to the number of nodes used for calibration. 
The method thus represents a valuable tool for gaining insights on privacy-protected systems. 
\end{abstract}
%\keywords{Complex Networks, Network Reconstruction, Exponential Random Graphs, Fitness model}

\section{Introduction}\label{intro}

The reconstruction of the statistical properties of a network when only limited information is available represents one of the outstanding and unsolved problems in the field of complex networks 
\cite{Clauset2008,Mastromatteo2012}. A first example is the case of financial networks, for which systemic risk estimation is based on the inter-dependencies among institutions 
\cite{Battiston2012a,Battiston2012b}---yet, due to confidentiality issues, the information that regulators are able to collect on mutual exposures is very limited \cite{Wells2004}. 
Other examples include social networks, for which information may be unavailable because of privacy problems or simply for the impossibility to sample the whole system. 

Network reconstruction has been typically pursued through Maximum Entropy (ME) algorithms \cite{Lelyveld2006,Degryse2007,Mistrulli2011}, which obtain link weights via a maximum homogeneity principle; 
however, the strong limitation of these algorithms resides in the assumption that the network is fully connected (for this reason they are known as ``dense reconstruction methods''), 
while real networks show a largely heterogeneous connectivity distribution. More refined methods like ``sparse reconstruction'' algorithms \cite{Mastromatteo2012} allow to obtain a network 
with arbitrary heterogeneity, but still cannot set an appropriate value for such heterogeneity. Recently, a novel {\em bootstrapping} (BS) method \cite{Musmeci2013,Caldarelli2013} has been proposed 
in order to overcome these problems. The BS method uses the limited information on the system to generate an ensemble of networks according to the {\em exponential random graph} (ERG) model 
\cite{Park2004}---where, however, the Lagrange multipliers that define it are replaced by {\em fitnesses}, \emph{i.e.}, known intrinsic node-specific properties related to the network topology \cite{Caldarelli2002}. 
The estimation of the network topological properties is then carried out within the ERG-induced ensemble. 
The method builds on previous results \cite{Garlaschelli2004} which showed that, in the particular case of the World Trade Web (see below), the knowledge of a non-topological property 
(the Gross Domestic Product), if coupled to that of the total number of links, allows to infer the topological properties of the network with great accuracy. 
This procedure can be restated within a maximum-likelihood framework \cite{Garlaschelli2008}. The BS method uses these preliminary observations 
to provide a reconstruction procedure valid in the case when the the degree sequence of the network (\emph{i.e.}, the number of connections for each node) is known only partially.

While past works \cite{Mastromatteo2012,Lelyveld2006,Degryse2007,Mistrulli2011,Musmeci2013} mainly dealt with using the limited information available on the network to estimate 
specific high-order properties such as systemic risk, in the present paper we employ the BS method to reconstruct the fundamental properties that are commonly used to describe a network: 
density of links, assortativity, clustering (see sec. \ref{sec:top_prop}). By focusing on these previously untested properties, we are able to enlarge the basket of quantities 
that are properly estimated by the BS approach. To validate our method we study how its accuracy depends upon the size of the subset of nodes for which the information is available; 
our case-study includes synthetic networks generated through a {\em fitness} model \cite{Caldarelli2002} as well as real instances of networked systems: 
1) the World Trade Web (WTW), \cite{Gleditsch2002}, \emph{i.e.}, the network whose nodes are the countries and links represent trade volumes among them, 
and 2) the network of interbank loans of the e-mid (E-mid) interbank money market \cite{DeMasi2006}.

\section{Method}

We start by briefly describing the ERG model and the fitness model, on which the BS method builds. 

The ERG model is one of the most common network generation framework \cite{Park2004,Dorogovtsev2010,Garlaschelli2009}, 
which consists in defining an ensemble $\Omega$ of networks which is maximally random, except for the ensemble average 
of a set of network properties $\{\langle C_a\rangle_\Omega\}$---constrained to some specific values $\{ C_a^* \}$. 
The probability distribution over $\Omega$ can then be defined via a set of control parameters $\{\theta_a\}$, namely the set of Lagrange multipliers associated with the constraints $\{C^*_a\}$.
A particular yet widely used case of the ERG model is known as the Configuration Model (CM) \cite{Park2004}, which is obtained by specifying the mean degree sequence $\{k_i^*\}_{i=1}^N$ of the network. 
In this case, each node $i$ is identified by the Lagrange multiplier $\theta_i$ associated to its degree $k_i$. 
By defining $x_i \equiv e^{-\theta_i}$ $\forall i$, the ensemble probability that any two nodes $i$ and $j$ are connected reads \cite{Park2004}:
\begin{equation}
 p_{ij}=\frac{x_ix_j}{1+x_ix_j}
\label{eq:prob linking}
\end{equation}
so that $x_i$ quantifies the ability of node $i$ to create links with other nodes (induced by its degree $k_i$) \cite{Squartini2011}.

On the other hand, the \textit{fitness} model \cite{Caldarelli2002} assumes the network topology to be determined by an intrinsic non-topological property (known as {\em fitness}) 
associated with each node of the network, and has successfully been used in the past to model several empirical economical networks \cite{Garlaschelli2005,DeMasi2006,Garlaschelli2004}. 
\newline

The BS method \cite{Musmeci2013} combines these two network generation models, working as follows. We start from incomplete information about the topology of a given network $G_0$ (consisting of $N$ nodes): 
we assume to know the degree sequence $\{k_i^*\}_{i\in I}$ of only a subset $I$ of the nodes (with $|I|=n<N$) and an intrinsic, non-topological property $\{y_i\}_{i\in V}$ for all the nodes---that will be our fitness 
(see below). Using this information, we want to find the most probable estimate of the value $X\,(G_0)$ of a topological property $X$ computed on the network $G_0$, 
compatible with the aforementioned constraints. The method builds on two important assumptions:
\begin{enumerate}
 \item The network $G_0$ is interpreted as drawn from an ERG-induced ensemble $\Omega$. We then expect the quantity $X(G_0)$ to mostly vary within the range $\langle X \rangle_{\Omega} \pm \sigma_X^{\Omega}$, 
 where $\langle X \rangle_{\Omega}$ and $\sigma_X^{\Omega}$ are respectively average and standard deviation of property $X$ estimated over the ensemble $\Omega$.
 \item The non-topological fitnesses $\{y_i\}$ are assumed to be proportional the degree-induced exponential Lagrange multipliers $\{x_i\}$ through a universal (unknown) parameter $z$:\footnote{Fitnesses 
 are often used within the ERG framework provided an assumed connection between them and the Lagrange multipliers. For instance, countries Gross Domestic Products (GDPs) work well as fitnesses 
 when modeling the WTW, and eq. (\ref{eq:prob linking}) accurately describes the WTW topology when $-\theta_i\propto \log(\mbox{GDP}_i)$ \cite{Garlaschelli2004}. 
 In any case, this second assumption (or any other relation $y_i=f(x_i)$) can be appropriately tested on the subset $I$ of nodes for which the degree is known.} $x_i\equiv\sqrt{z}y_i$ $\forall i$. 
Therefore eq. (\ref{eq:prob linking}) becomes: 
\begin{equation}
 p_{ij}=\frac{zy_iy_j}{1+zy_iy_j}. \label{eq:prob linking z}
\end{equation}
\end{enumerate}
Thanks to these two assumptions we can turn the problem of evaluating $X\,(G_0)$ into the one of choosing the optimal ERG ensemble $\Omega$ compatible with the constraints on $G_0$, 
which is the most appropriate to extract the real network $G_0$ from---given that we  know only partial information. 
Once $\Omega$ is determined (by the set $\{x_i\}$ and thus by the set $\{y_i\}$), we can use the average $\langle X \rangle_{\Omega}$ as a good estimation for $X\,(G_0)$ 
and $\sigma_X^{\Omega}$ as the typical statistical error. Now, since we know the rescaled fitness values $\{y_i\}$, the problem becomes equivalent 
to that of finding the most likely value of $z$ that defines the ensemble $\Omega$ according to eq.~(\ref{eq:prob linking z}).

An estimation for the value of $z$ can be found using the incomplete degree sequence through the following relation \cite{Musmeci2013}:
\begin{equation}\label{eq:estimatez}
\sum_{i \in I} \langle k_i \rangle_\Omega \equiv \sum_{i \in I} \sum_{\substack{j=1\\j \neq i}}^N p_{ij} = \sum_{i \in I} k_i^*
\end{equation}
in which the first equality comes from the definition of the ERG model, and the second one is the application of the maximum-likelihood argument \cite{Park2004}---the whole equation being restricted 
only to the nodes belonging to the subset $I$. 
Since $\langle k_i\rangle_{\Omega}=\sum_{j(\neq i)}p_{ij}$ contains the unknown $z$ through eq. (\ref{eq:prob linking z}), and $\{y_i\}$ and $\sum_{i \in I} k_i^*$ are known, eq.~(\ref{eq:estimatez}) 
defines an algebraic equation in $z$, whose solution can be used to estimate the degree-induced Lagrange multipliers $x_i=\sqrt{z}y_i$ $\forall i$ of the ERG ensemble $\Omega(z)$, 
and at the end to obtain an estimation of $X\,(G_0)$. 

Summing up, the BS algorithm consists in the following steps. Given a network $G_0$, the knowledge of some non-topological property $\{y_i\}$ for all the nodes 
and the knowledge of the degrees of a subset $I$ of nodes:
\begin{itemize}
\item compute the sum of the degrees of the nodes in $I$ ($\sum_{i \in I} k_i^*$) and use it together with $x_i=\sqrt{z}y_i$ $\forall i$ to solve eq. (\ref{eq:estimatez}) and to obtain the corresponding value of $z$;
\item using the estimated $z$ and the knowledge of $\{y_i\}$, generate the ensemble $\Omega(z)$ by placing a link between each pair of nodes $i$ and $j$ with probability $p_{ij}$ 
given by eq.~(\ref{eq:prob linking z});
\item compute the estimate of $X\,(G_0)$ as $\langle X \rangle_{\Omega}\pm\sigma_X^{\Omega}$, either analytically or numerically.
\end{itemize}

\section{Topological Properties}\label{sec:top_prop}

As stated in the introduction, in testing the BS method we will focus on the topological properties (each playing the role of $X$ in the previous discussion) 
which are commonly regarded as the most significant for describing a network. To define these properties, we use the formalism of the adjacency matrix---where $a_{ij}=1$ if nodes $i$ and $j$ 
are connected, and $a_{ij}=0$ otherwise. We consider:
\begin{itemize}
\item link density (or connectance):
\begin{equation}\label{eq.D}
D:=\frac{\sum_{i<j}a_{ij}}{N(N-1)/2}=\frac{2L}{N(N-1)}
\end{equation}
which is the ratio between the actual number of links in the network and the maximal one compatible with the number of nodes $N$;
\item mean average nearest-neighbors degree:
\begin{equation}\label{eq.k_nn}
k_{nn}:=\frac{\sum_i k_{nn,\:i}}{N}\mbox{~~where~~}k_{nn,\:i}:=\frac{\sum_{j(\neq i)}a_{ij}k_{j}}{k_i}=\frac{\sum_{j(\neq i)}\sum_{k(\neq i,j)}a_{ij}a_{jk}}{\sum_{j(\neq i)}a_{ij}}
\end{equation}
\emph{i.e.}, the arithmetic mean of the degrees of the neighbors of each node, averaged over all nodes;
\item the mean clustering coefficient:
\begin{equation}\label{eq.clu}
c:=\frac{\sum_i c_{i}}{N}\mbox{~~where~~}c_i:=\frac{\sum_{j(\neq i)}\sum_{k(\neq i,j)}a_{ij}a_{ik}a_{jk}}{\sum_{j(\neq i)}\sum_{k(\neq i,j)}a_{ij}a_{ik}}
\end{equation}
\emph{i.e.}, the ratio between the number of observed links in each node neighborhood and the maximum possible number of such links, averaged over all nodes \cite{Watts1998};
\item average rich-club coefficient:
\begin{equation}\label{eq.phi}
\phi=\sum_k\,P(k)\,\phi(k)\mbox{~~where~~}\phi(k)=\frac{\psi-D}{1-D}\mbox{~~and~~}\psi(k)=\frac{2E_{>k}}{N_{>k}(N_{>k}-1)}
\end{equation}
with $P(k)$ representing the fraction of nodes having degree equal to $k$ and $\phi(k)$ the ratio between the $E_{>k}=\sum_{i:k_i>k}\sum_{j:k_j>k}a_{ij}$ edges 
actually connecting the $N_{>k}=\sum_i\Theta\left[\sum_{j}a_{ij}-k\right]$ nodes with degree higher than $k$ 
and the maximum possible number $N_{>k}(N_{>k}-1)/2$ of such edges (or, in other words, the density of links of the subgraph consisting of only the nodes with degree higher than $k$) \cite{Colizza2006}.
\end{itemize}

\section{Dataset}\label{sec:data}

In order to validate our BS method we use two instances of real economic systems. The first one is the World Trade Web \cite{Gleditsch2002}, \emph{i.e.}, the network whose nodes represent the world-countries 
and links represent trade volumes between them:\footnote{For WTW, we use trade volume data for the year 2000.} thus, $w_{ij}$ is the total amount of the export of country $i$ to country $j$. 
The second one is the interbank loan network of the so-called E-mid interbank money market \cite{DeMasi2006}. In this case, the nodes represent the banks and a link $w_{ij}$ between banks $i$ and $j$ 
represents the amount of the loan from bank $i$ to bank $j$.\footnote{For E-mid, we consider snapshots of loans aggregated on a monthly scale (as also done in other works \cite{DeMasi2006}) because of the high volatility of the links 
at shorter time scales. In the following, we report the results about the snapshot for February 1999. We performed the same analysis also for other monthly snapshots and we found comparable results.} 
These datasets are particularly suited for our study, as the node fitnesses $\{y_i\}$ can be naturally identified with country GDP 
for the WTW and with the banks total exposures (\emph{i.e.}, the total lending) for E-mid. Thus in both cases each node fitness coincides with its total strength: $y_i\equiv s_i=\sum_j w_{ij}$. 
The binary undirected version of these networks (that we want to reconstruct) is then built as $a_{ij}=\Theta[w_{ij}+w_{ji}]$.

\section{Test of the BS method}\label{sec:test-bm_binary}

\begin{figure}[t!]
\begin{center}
\includegraphics[width=0.7\textwidth]{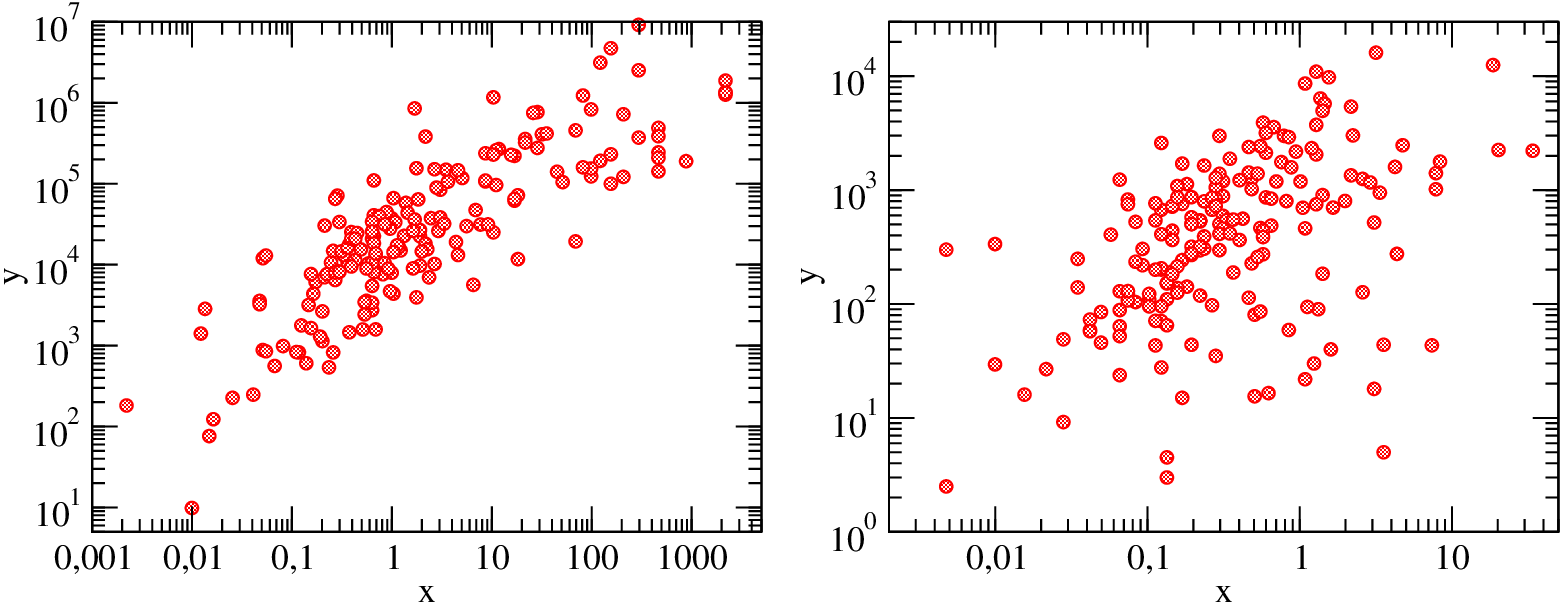}
\end{center}
\caption{Scatter plot of node fitness $\{y_i\}$ vs exponential Lagrange multiplier $\{x_i\}$ for WTW (left panel) and E-mid (right panel).}\label{fig:test_fit}
\end{figure}

Before proceeding to results, we remark that the BS method is subject to two different types of errors. The first one is due to the limited information 
available for calibrating the ERG model: since we know only the degrees of a subset $I$ of nodes, we can just obtain an estimate of the best $z$ of the ERG ensemble through eq. (\ref{eq:estimatez}). 
The second error comes from the assumption that the node fitnesses $\{y_i\}$ are proportional to the degree-induced Lagrange multipliers $\{x_i\}$. 
Fig. \ref{fig:test_fit} shows the relation between $\{y_i\}$ and $\{x_i\}$ for the two empirical networks. Indeed, there are deviations 
from linearity---which would correspond to a perfect realization of the fitness model. Note that a better correlation is observed for the WTW: thus, we can expect the BS method to work better in this case. 
\newline\newline
The quantitative estimation of the BS method effectiveness in reconstructing the topological properties of the two case-studies networks thus proceeds in two steps, as follows. 

\paragraph{Test on synthetic networks.} To assess the errors which are only due to the limited information available about the degree sequence, we first perform a benchmark test on 
{\em synthetic} networks generated through the fitness model. This means that we use the fitnesses $\{y_i\}$ (GDPs for the WTW and total loans for E-mid) to evaluate the ``real'' $z$ 
of the ERG ensemble by solving eq. (\ref{eq:estimatez}) with all nodes included in $I$ (\emph{i.e.}, assuming to know the whole degree sequence), and then draw a network $G_0$ from $\Omega(z)$ 
by numerically generating it through eq. (\ref{eq:prob linking z}). 
$G_0$ is now the network to reconstruct through the BS method (\emph{i.e.}, with partial information), and the value of $X$ is computed both on $G_0$ itself as $X(G_0)$, as well as on the whole ensemble 
$\Omega(z)$ as $\langle X\rangle_{\Omega(z)}$.\footnote{In the latter case, we use $p_{ij}=\langle a_{ij}\rangle_{\Omega(z)}$ as the expected values of the adjacency matrix elements $a_{ij}$ in the 
definitions (\ref{eq.D},\ref{eq.k_nn},\ref{eq.clu},\ref{eq.phi}).} Inferring $G_0$ then consists of the following operative steps:
\begin{itemize}
\item choose a value of $n<N$ (the number of nodes for which the degree is known);
\item build a set of $M=100$ subsets $\{I_\alpha\}_{\alpha=1}^M$ of $n$ nodes picked at random;
\item for each subset $I_\alpha$, use the degree sequence in $I_\alpha$ to evaluate $z_\alpha$ from eq. (\ref{eq:estimatez}), and use it to build the ensemble $\Omega(z_\alpha)$; 
\item use the linking probabilities from eq. (\ref{eq:prob linking z}) to compute the value of property $X$ over the ensemble $\Omega(z_\alpha)$ as $X_\alpha=\langle X \rangle_{\Omega(z_\alpha)}$;
\item compute the relative root mean square error (rRMSE) of property $X$ over the subsets $\{I_\alpha\}$:
\begin{equation}\label{eq.RMSE}
r_X\equiv\sqrt{\frac{1}{M}\sum_{\alpha=1}^M\left[\frac{X_\alpha}{X_0}-1\right]^2}
\end{equation}
\end{itemize}
In the rRMSE expression, $X_0$ denotes a reference value of property $X$, which can be either $X(G_0)$ (the value of $X$ measured on $G_0$), as well as $\langle X\rangle_{\Omega(z)}$ 
(the value of $X$ on the whole ensemble $\Omega(z)$). The consequent two alternative rRMSEs (which we denote as $r_X^0$ and $r_X^{\Omega_0}$, respectively) 
provide different estimates of the BS method accuracy. In fact, $r_X^0$ tests the ability to reproduce a single outcome of the ensemble sampling, whereas, $r_X^{\Omega_0}$ 
refers to the theoretical values of $X$ expected on the ensemble. Note that since the ensemble $\Omega(z)$ is generated through the fitness model, by construction $r_X^{\Omega_0}\rightarrow0$ 
for $n\rightarrow N$; however, $r_X^0$ does not necessarily tend to zero, because the generated configuration $G_0$ is a single realization of $\Omega(z)$ and thus in general $X(G_0)\neq\langle X\rangle_{\Omega(z)}$. 

\begin{figure}[t!]
\begin{center}
\includegraphics[width=0.8\textwidth]{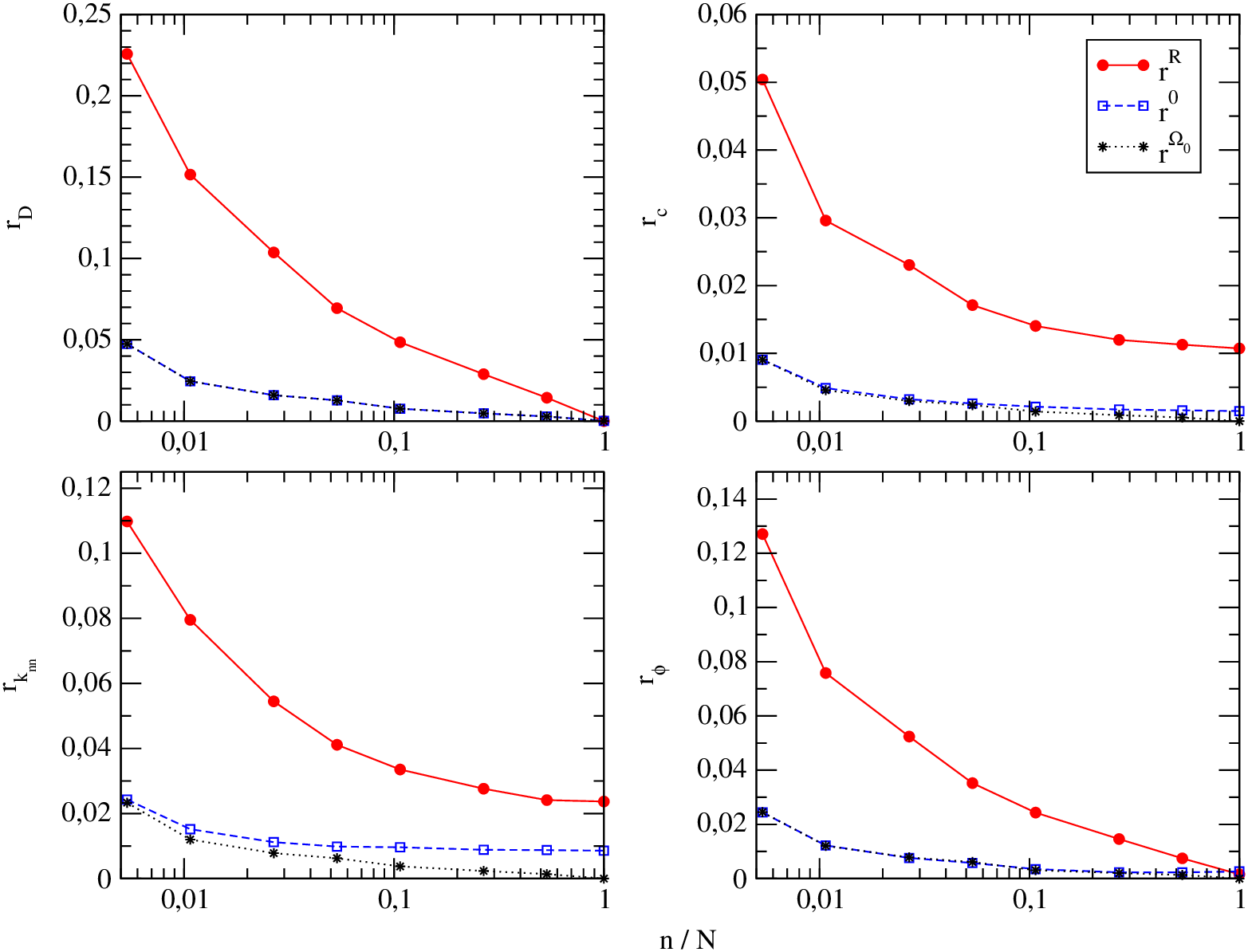}
\end{center}
\caption{rRMSE of various topological properties for different values of $n$, obtained by the BS method on the real WTW network and on its synthetic version obtained through the fitness model.
Top left: density $D$; top right: clustering coefficient $c$; bottom left: average nearest-neighbor degree $k_{nn}$; bottom right: rich-club coefficient $\phi$.}
\label{fig:test_on_WTW}
\end{figure}

\paragraph{Test on real networks.} The testing procedure for real networks is equivalent to the one described above, with the only difference that $G_0$ and $X_0$ are now the empirical network 
and the value of $X$ computed on such network respectively---the rRMSE in this case is denoted as $r_X^R$. We recall again that since the fitness model is only an approximation for real networks, 
in this case we expect larger errors from the BS reconstruction than those computed for the synthetic networks. 

\begin{figure}[t!]
\begin{center}
\includegraphics[width=0.8\textwidth]{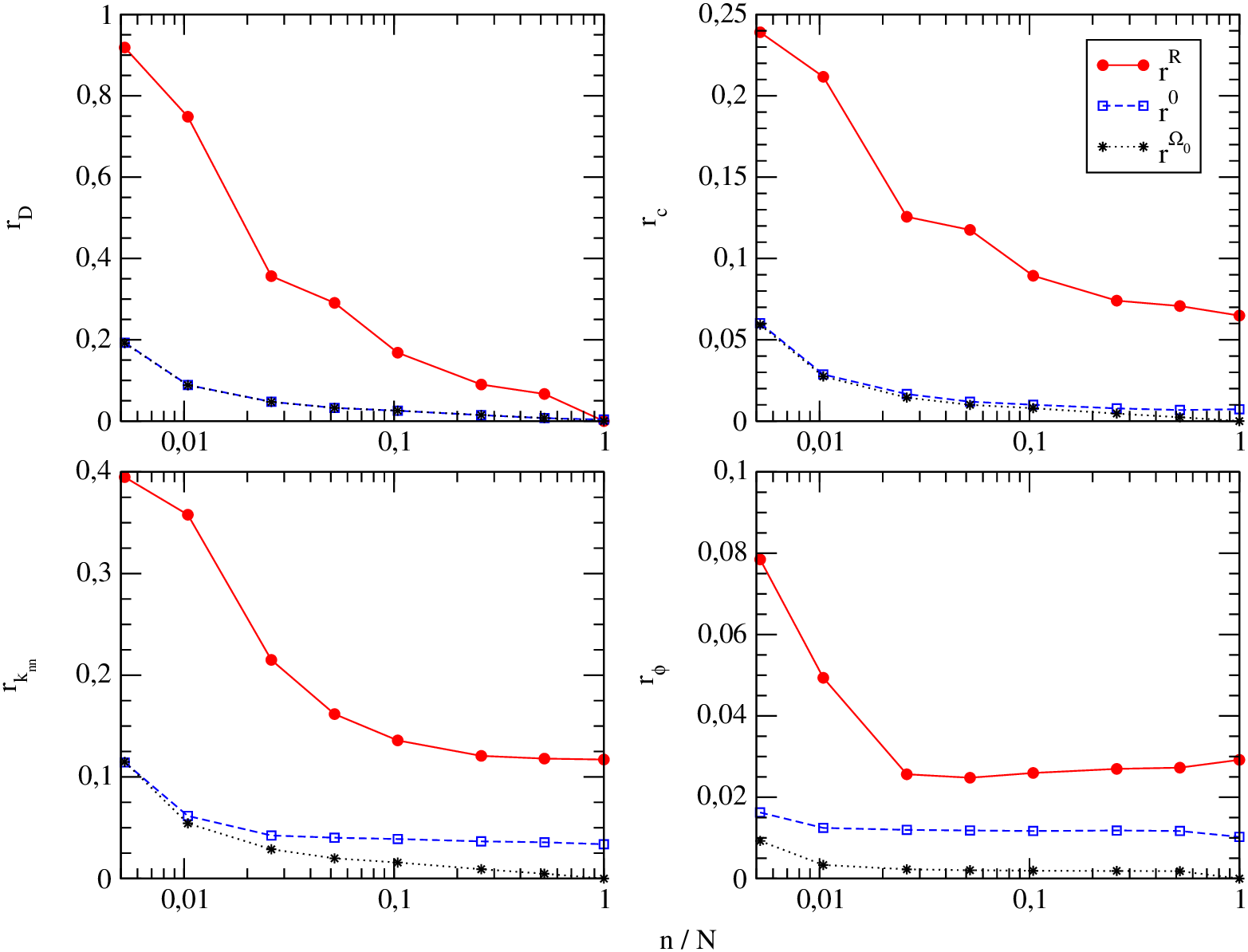}
\end{center}
\caption{rRMSE of various topological properties for different values of $n$, obtained by the BS method on the real E-mid network and on its synthetic version obtained through the fitness model. 
Top left: density $D$; top right: clustering coefficient $c$; bottom left: average nearest-neighbor degree $k_{nn}$; bottom right: rich-club coefficient $\phi$.}
\label{fig:test_on_emid}
\end{figure}

\paragraph{Results.} In order to study the accuracy of the BS reconstruction, we study how the rRMSE for the various topological properties we consider varies as a function of the size $n$ of the subset of nodes 
used to calibrate the ERG model (\emph{i.e.}, for which information about the degrees is available). Results are shown in Fig.\ref{fig:test_on_WTW} for WTW and in Fig. \ref{fig:test_on_emid} for E-mid. 
We observe that in all cases there is a rapid decrease of the relative error as the number of nodes $n$, used to reconstruct the topology, increases. 
This is an indication of the goodness of the estimation provided by the BS method. As expected, the rRMSE is higher for real networks than for synthetic networks, 
and the difference between the two respective curves gives a quantitative estimation of the error made in modeling real networks with the fitness model. 
The fact that such difference is higher for E-mid than for WTW is directly related to the better correlation between node fitnesses and node degrees observed in the latter case. 
Note also that the errors for E-mid are higher than the ones for WTW also for the corresponding synthetic networks. 
This feature is easily explained by the higher link density of WTW ($D\simeq0.59$) with respect to that of E-mid ($D\simeq0.20$): 
denser networks are easier to reconstruct because nodes have more links, and thus carry more information for the BS method to exploit (as fluctuations are tamed out).

\section{Conclusions}\label{sec:conclusions}

In this paper we tested a novel network reconstruction (BS) method that allows to estimates the topological properties of a network by using only partial information about its connectivity, 
as well as a non-topological quantity (interpreted as fitness) associated to each node. This method is particularly useful to overcome the lack of topological information 
which often hinders the study of complex networks. We tested the method on empirical networks, as well as on synthetic networks generated through the fitness model, 
and studied how well it can estimate the fundamental topological properties which are widely employed to describe network patterns: connectivity, assortativity, clustering coefficient and rich-club coefficient. 
We found that these properties are reconstructed accurately, for instance with a tolerance usually varying from 2\% to 15\% (depending on the property examined) using only 5\% of the nodes. 
We also found that the BS method brings to better estimates in denser networks (where more information is available); additionally, the method effectiveness strongly depends on the accuracy of the fitness model 
used to describe the empirical dataset. In the case of the WTW, the fitness model is fairly accurate in describing how links are formed across countries depending on their GDP \cite{Garlaschelli2004} 
and the BS is thus efficient in reconstructing the network topological properties. In the case of E-mid, the fitness model is less accurate and so is the BS method, but the latter can still lead to useful outcomes.

While at first thought it can be surprising that a small fraction of nodes enables to estimate with high accuracy global emerging properties of the network, 
it is important to remark that the BS method assumes the knowledge of the fitness of all nodes and the validity of the fitness model in describing the data. 
Therefore, a limitation of this method could arise when considering higher-order topological properties as the community structure. 
Possibly, in these situations the method could require a larger initial information to obtain the same results. Investigation of these cases is left for future research. 
Finally note that, contrarily to past works \cite{Mastromatteo2012,Lelyveld2006,Degryse2007,Mistrulli2011,Musmeci2013}, in this paper we focused on estimating previously untested network quantities, 
namely the most fundamental properties of a network. This allows us to enlarge the number of properties whose estimations are in remarkable agreement with observations. 
Indeed, any network reconstruction method must be tested against these basic quantities before being employed for more demanding tasks that are specific for a class of networks (such as systemic risk estimation 
in financial networks). Thus, the validation of the BS method we presented here also allows us to extend its applicability to any set of dependencies among components in a complex system.

\section*{Acknowledgments}
This work was supported by the EU project GROWTHCOM (611272), the Italian PNR project CRISIS-Lab, the EU project MULTIPLEX (317532) and the Netherlands Organization for Scientific Research (NWO/OCW).
DG acknowledges support from the Dutch Econophysics Foundation (Stichting Econophysics, Leiden, the Netherlands) with funds from beneficiaries of Duyfken Trading Knowledge BV (Amsterdam, the Netherlands).

\end{document}